\pdfoutput = 1 

\documentclass[aps,prd,twocolumn,nofootinbib,superscriptaddress,floatfix,showpacs,showkeys,
preprintnumbers,
  amsmath,
  amssymb,
  aps,
  prd
]{revtex4-2}

\usepackage{cancel}
\usepackage{multirow}

\usepackage{graphicx}
\usepackage{dcolumn}
\usepackage{bm}

\usepackage[colorlinks=true,
linkcolor=purple,
anchorcolor=magenta,
citecolor=blue
]{hyperref}

\usepackage[english]{babel}
\usepackage{xcolor}
\usepackage{mathtools}
\usepackage{pdfsync}
\usepackage{amssymb}
\usepackage{amsmath}

\usepackage{slashed}

\usepackage{marvosym}

\usepackage[squaren]{SIunits}

\usepackage[active]{srcltx}

\def \beq  {\begin{equation}}
\def \eeq  {\end{equation}}

\newcommand{\overbar}[1]{\mkern 1.5mu\overline{\mkern-1.5mu#1\mkern-1.5mu}\mkern 1.5mu}
\newcommand{\MSbar}{\overbar{\text{MS}}}

\def\II{\hbox{{1}\kern-.25em\hbox{l}}}

\newcommand \widebar [1] {\overline{#1}}

\newcommand{\PS}{\text{\tiny PS}}
\newcommand{\NS}{\text{\tiny NS}}
\newcommand{\CF}{\text{\tiny F}}
\newcommand{\CA}{\text{\tiny A}}
\newcommand{\NP}{\text{\tiny NP}}

\usepackage{array}   
\newcolumntype{L}{>{$}c<{$}} 

\begin{document}
\allowdisplaybreaks

\preprint{DESY-25-187}


\title{Conformal moments of the  two-loop coefficient functions in DVCS}

\author{Vladimir~M.~\surname{Braun}}

\affiliation{Institut f{\"u}r Theoretische Physik, Universit{\"a}t Regensburg, D-93040 Regensburg, Germany}

\author{Patrick~\surname{Gotzler}}

\affiliation{Institut f{\"u}r Theoretische Physik, Universit{\"a}t Regensburg, D-93040 Regensburg, Germany}

\author{Alexander~N.~\surname{Manashov}}

\affiliation{II.~Institut f\"ur Theoretische Physik, Universit\"at Hamburg, D-22761 Hamburg, Germany}

\affiliation{Institut f{\"u}r Theoretische Physik, Universit{\"a}t Regensburg, D-93040 Regensburg, Germany}

\begin{abstract}%
We develop a new technique and calculate conformal (Gegenbauer) moments of the two-loop
coefficient functions in Deeply Virtual Compton Scattering (DVCS). These results 
are necessary for the extraction of the generalized parton distributions from the 
experimental data to the NNLO accuracy within the Mellin-Barnes approach.  
\end{abstract}

\date{\today}

 \maketitle
\section{Introduction}\label{sect:introduction}%

The multidimensional composition of the nucleon in terms of quarks and gluons
has drawn growing interest in the past decades and became 
an important frontier for the study of strong interactions. 
Generalized parton distributions (GPDs) \cite{Muller:1994ses} encode the 3-dimensional
distributions of partons in the momentum fraction and the impact parameter space.
Their study has been one of the main physics motivations for the 
future upgrades at JLAB \cite{Accardi:2023chb}, 
the Electron-Ion Collider (EIC) \cite{Accardi:2012qut} in U.S.A. and the  EIcC project 
\cite{Anderle:2021wcy} in China.

The Deep Inelastic Compton Scattering (DVCS) \cite{Ji:1996ek,Radyushkin:1996nd} plays an exceptional role in this endeavor.
This is the simplest process that involves GPDs. The underlying formalism is well understood 
\cite{Radyushkin:1997ki,Ji:1998xh,Collins:1998be} and in the last years considerable effort was invested to advance 
the theory description to the NNLO accuracy. The two-loop DVCS coefficient functions (CFs) were calculated in 
Refs.~\cite{Braun:2020yib,Braun:2021grd,Gao:2021iqq,Braun:2022bpn,Ji:2023xzk} for all cases, and the three-loop
evolution equation for the flavor-nonsinglet GPDs was derived in Refs.~\cite{Braun:2017cih,Ji:2023eni}.
The flavor-singlet three-loop evolution equation is not known yet. However, the phenomenological impact of the 
renormalization group improvement at three loops is expected to be minor since the range of photon virtualities
accessible in current and future DVCS experiments is rather limited. 

Implementation of these new theory results in the data analysis aiming at the  GPD extraction 
is by itself a nontrivial problem. 
It has been suggested and extensively discussed in the literature
\cite{Manashov:2005xp,Mueller:2005ed,Kirch:2005tt,Kumericki:2007sa,Kumericki:2009uq,Zhang:2024djl}
that this analysis can be facilitated by going over from the momentum fractions to the conformal moments 
space and evaluating the convolutions using the Mellin-Barnes techniques.
The original motivation to use this representation was that in this way the one-loop evolution equations
become diagonalized and can be solved easily. A fast NLO evolution code using Mellin-Barnes 
formalism also exists \cite{Gepard}. It was realized that going over to conformal moments
is also beneficial in another respect: it allows one to utilize various physical constraints on the GPDs, 
and provides one with a natural platform to construct GPD parametrizations.   
In particular, the Kumeri\v{c}ki-M{\"u}ller (KM) model
was proposed and used successfully in fitting the DVCS
and deeply virtual meson production (DVMP) measurements
\cite{Kumericki:2009uq,Muller:2013jur,Cuic:2023mki}. 
The same framework is employed in the GUMP initiative 
\cite{Guo:2022upw,GUMP,Guo:2024wxy,Guo:2025muf} which aims at the global GPD extraction 
analysis from extensive experimental inputs, see also \cite{Mamo:2024jwp,Mamo:2024vjh,Hechenberger:2025wnz}.

Conformal moments of the DVCS CFs that are needed in this approach are given by 
(analytically continued) integrals of the momentum fraction space CFs with Gegenbauer polynomials, 
e.g. for quarks 
\begin{align}
  C_q(j) \sim \int_{-1}^1\!dx\, (1-x^2) C_q(x)\, C^{3/2}_j(x)\,.
\label{scheme1}
\end{align}
Such integrals become nontrivial in higher orders,
as the momentum fraction space CFs are given in terms of generalized polylogarithms of increasing transcendentality. 
In this paper we present a systematic approach how the Gegenbauer moments can be evaluated with moderate 
computational effort, and present explicit expressions for the conformal moments of all two-loop DVCS 
CFs~\cite{Braun:2020yib,Braun:2021grd,Gao:2021iqq,Braun:2022bpn,Ji:2023xzk} --- vector and axial vector, 
flavor-singlet and flavor nonsinglet, and also for gluon transversity.

To explain the basic idea on an intuitive level, it is convenient to use 
Dirac bra-ket notation  $C^{3/2}_j \leftrightarrow |j\rangle$. 
The conformal moments \eqref{scheme1} can be thought of as scalar products 
$\langle C_q |j\rangle$. 
The main observation is that such ``states'' (Gegenbauer polynomials) 
are eigenfunctions of $\mathrm{SL}(2)$-invariant operators $\mathbb H_\alpha$, schematically
\begin{align}
   \mathbb H_\alpha |j\rangle  = E_\alpha(j) |j\rangle\,,  
\end{align}
where $\alpha$ is a certain index used to enumerate such operators.
For a (more or less) arbitrary function $f(x)$
obviously  
\begin{align}
 \langle \mathbb H_\alpha^\dagger f|j\rangle = \langle f|\mathbb H_\alpha|j\rangle = E_\alpha(j)\langle f |j\rangle\,, 
\end{align} 
so that if $\langle f |j\rangle$ (conformal moment of the function $f(x)$) is known and the eigenvalue
$E_\alpha(j)$ can be calculated, one obtains from this relation the conformal moment of a (potentially more complicated)
function $\mathbb H_\alpha^\dagger f$. 
Using such relations for a sufficiently large set of independent $\mathrm{SL}(2)$-invariant operators 
(at two loops we will need sixteen operators) one can calculate conformal moments for a 
complete basis of functions with a given weight, from which the results for the DVCS CFs are obtained 
by solving a system of linear equations.
An important technical issue that will be explained in what follows is that it proves to be sufficient to use 
position space expressions for the invariant operators, which have a simple form. 
The calculation was done using the \texttt{HyperInt} package~\cite{Panzer:2014caa} and we also used
\texttt{PolyLogTools} \cite{Duhr:2019tlz} to simplify the final expressions.

\section{Definitions and notations}\label{sect:definitions}%

The DVCS amplitudes $\gamma^\ast(q) + N(p) \to \gamma(q')+N(p')$ 
are defined by the nucleon matrix element of the time-ordered product of two electromagnetic currents
and can be split in the vector $V$ and axial-vector $A$ parts that are symmetric and antisymmetric in the Lorenz indices,
respectively: 
\begin{align}
T_{\mu\nu}  & = i\int d^4 x\, e^{-iqx} \langle p'| T\{ j^\mu(x)j^\nu(0)\}|p\rangle\,
\notag\\
&= -g_{\mu\nu}^\perp V +\epsilon_{\mu\nu}^\perp A +\text{power corrections}.
\end{align}
The amplitudes $V$ and $A$, in turn, 
are usually expressed in terms of the so-called Compton Form Factors (CFFs), see e.g. Ref.~\cite{Belitsky:2005qn},
\begin{align}\label{VAdef}
V&=\frac{1}{2P_+} \bar u(p')\left[\gamma^+ \mathcal H(\xi,Q,t) +\frac{i\sigma^{+\alpha}\Delta_\alpha}{2M}\mathcal E(\xi,Q,t)\right] u(p),
\notag\\[2mm]
A&=\frac{1}{2P_+} \bar u(p')\left[\gamma^+ \widetilde{\mathcal H}(\xi,Q,t)
 +\frac{\Delta_+}{2M}\widetilde{\mathcal E}(\xi,Q,t)\right]\gamma_5 u(p),
\end{align}
where $P=(p+p^\prime)/2$, $\Delta=q'-q$, $Q^2=-q^2$, and $M$ is the nucleon mass. The skewedness parameter $\xi$ can be
taken as 
\begin{align}
\xi=\frac{x_B}{2-x_B} +\mathcal O(1/Q^2),
\end{align}
where $x_B=Q^2/(2pq)$ is the Bjorken variable.

To the leading twist accuracy, 
the  CFFs $\mathcal H,\mathcal E,\widetilde{\mathcal H},\widetilde{\mathcal E}$ can be written  in the factorized form
~\cite{Radyushkin:1997ki,Ji:1998xh,Collins:1998be}, e.g., 
\begin{align}\label{Hfactorform}
\mathcal H&= \sum_q \int_{-1}^1 \frac{dx}\xi C_q\left(x/\xi,\mu^2/Q^2,\alpha_s(\mu)\right) H_q(x,\xi,t,\mu)
\notag\\
&\quad
+\int_{-1}^1 \frac{dx}{\xi^2} C_g\left(x/\xi,\mu^2/Q^2,\alpha_s(\mu)\right) H_g(x,\xi,t,\mu),
\end{align}
where $H_q,\, H_g$ are the quark (gluon) GPDs and $C_q,\, C_g$ are the corresponding CFs, and similarly for the other CFFs. 
The CFs for the vector and axial-vector cases are different --- we will use the notation $\widetilde C_q\,, \widetilde C_g$
for the axial vector ones --- but are the same for the pairs $\mathcal H$, $\mathcal E$  and  
$\widetilde{\mathcal H}$, $\widetilde{\mathcal E}$, respectively.
The CFs $C_{q,g} (x/\xi)$,  $\widetilde C_{q,g} (x/\xi)$ are real functions at $|x/\xi|<1$ and can be continued 
analytically beyond this region using the $\xi\to\xi-i\epsilon$ prescription~\cite{Ji:1996ek,Radyushkin:1996nd}. 
The CFs $C_q$  and $\widetilde C_g$  are odd functions of $x$, while $C_g$ and $\widetilde C_q$ are even functions.

It has been suggested and extensively discussed in the literature
\cite{Manashov:2005xp,Mueller:2005ed,Kirch:2005tt,Kumericki:2007sa,Kumericki:2009uq,Zhang:2024djl}
that going over to the Mellin representation in convolution integrals in~\eqref{Hfactorform} can
have advantages for the data analysis (see Introduction). These integrals can be brought to the form, 
schematically
\begin{align}
\mathcal F&=\int_C \frac{dj}{2\xi}
\frac{(i\xi)^{-j}}{\cos\frac{\pi j}2}
\mathbb{C}^{(F)}_j\left({Q^2}/{\mu^2},a_s(\mu)\right) F_j(\xi,t,\mu)
\label{Mellin}
\end{align}
where $\mathcal F \in \{\mathcal H,\mathcal E,\widetilde{\mathcal H},\widetilde{\mathcal E}\}$,
$F_j$ are the conformal moments of the corresponding GPDs, and $\mathbb{C}^{(F)}_j$ are conformal moments
of the CFs which are given, up to an overall normalization, by the analytically continued integrals of 
the momentum-space CFs with Gegenbauer polynomials
\begin{align}
C_q(N) &= \frac14\int_{-1}^1 \!dx\, (1-x^2)\, C^{(3/2)}_{N-1} (x) \, C_q(x)\,,
\notag\\
C_g(N) &= \frac3{16}\int_{-1}^1 \!dx\, (1-x^2)^2\, C^{(5/2)}_{N-2} (x) \, C_g(x)
\label{GegMoments}
\end{align}
and similar for $\widetilde C^{q}_j$, $\widetilde C^{g}_j$. Here $N=j-1$ is the Lorentz spin of the corresponding
local operator.
The integral in \eqref{Mellin} is taken along a line parallel to the imaginary axis, such that all
singularities of the integrand are located to the left of this line. 
A thorough discussion and technical details can be found in
Refs.~\cite{Mueller:2005nz,Kumericki:2006xx,Kumericki:2007sa,Kumericki:2009uq,Guo:2024wxy,Zhang:2024djl}.

The CFs have a regular perturbative expansion
\begin{align}\label{CqCg}
C_q &=C_q^{(0)} +a_s C_q^{(1)}+a_s^2 C_q^{(2)} +\mathcal O(\alpha_s^3),
\notag\\
C_g & =a_s C_g^{(1)}+a_s^2 C_g^{(2)} +\mathcal O(\alpha_s^3),
\end{align}
where $a_s=\alpha_s/4\pi$, and similar for $\widetilde{ C}_q, \widetilde{ C}_g$.
The tree-level quark CFs take the form \cite{Ji:1996ek,Radyushkin:1996nd} 
\begin{align}
C_q^{(0)}(x)&=e_q^2\left(\frac{1}{1-x}-\frac1{1+x}\right),
\notag\\
\widetilde C_q^{(0)}(x)&=e_q^2\left(\frac{1}{1-x}+\frac1{1+x}\right).
\end{align}
The one-loop CFs $C_q^{(1)}, C_g^{(1)}, \widetilde C_q^{(1)}, \widetilde C_g^{(1)}$ in the $\MSbar$ scheme have  been  known
for a long time~\cite{Ji:1997nk,Mankiewicz:1997bk,Belitsky:1997rh,Ji:1998xh,Hoodbhoy:1998vm,Belitsky:2000jk} 
and the two-loop CFs have been calculated 
recently~\cite{Braun:2020yib,Braun:2021grd,Gao:2021iqq,Braun:2022bpn,Ji:2023xzk}. 
Explicit expressions for all CFs in momentum fraction space are collected in Refs.~\cite{Braun:2022bpn,Ji:2023xzk}.
However, the results for the flavor-nonsinglet axial-vector CF quoted in \cite{Ji:2023xzk} 
correspond to the Larin's renormalization scheme. For this case, we use instead the expressions 
from~\cite{Braun:2021grd,Gao:2021iqq} in the $\MSbar$ scheme. In this scheme, by definition, the anomalous
dimensions of flavor-nonsinglet vector and axial-vector operators coincide to all orders.   

As the final remark, the CFs in~\cite{Braun:2020yib,Braun:2021grd,Gao:2021iqq,Braun:2022bpn,Ji:2023xzk}
are written as functions of the variable
\begin{align}
z = (1-x)/2\,, && \bar z =1-z = (1+x)/2\,,
\end{align} 
which is more convenient for the calculations.
In terms of this variable, the tree level CFs are given by
\begin{align}
C_q^{(0)}(z)&= \frac{e_q^2}2 \left(\frac{1}{z}-\frac1{\bar z}\right),
\notag\\
\widetilde C_q^{(0)}(z)&= \frac{e_q^2}2 \left(\frac{1}{z}+\frac1{\bar z}\right),
\end{align}
and the Gegenbauer moments \eqref{GegMoments} take the form
\begin{align}
C_q(N) &=  2 \int_{0}^1 \!dz\, z\bar z\, C^{(3/2)}_{N-1} (1-2z) \, C_q(z)\,,
\notag\\
C_g(N) &=  6 \int_{0}^1 \!dz\, z^2\bar z^2\, C^{(5/2)}_{N-2} (1-2z) \, C_g(z)\,.
\label{GegMoments2}
\end{align}
%

\section{Kernels \& Moments }\label{sect:kernel+moment}%

In our normalization, Eq.~\eqref{GegMoments2}, moments of the tree-level vector and axial-vector quark CFs are
equal to 
\begin{align} 
 C_q(N) &= \frac{1+(-1)^N}{2} e_q^2 \,,\qquad \widetilde C_q(N) &= \frac{1-(-1)^N}{2} e_q^2 \,. 
\label{LO}
\end{align} 
In what follows we will tacitly assume that the expressions for vector and axial-vector CFs are written for 
even and odd N, respectively, and suppress $(1\pm(-1)^N)/2$ factors.

Beyond tree-level, the CFs are given by linear combinations of the harmonic 
polylogarithms (HPLs)~\cite{Remiddi:1999ew} with simple prefactors, 
\begin{align}\label{HPLa}
z^{-a} \, \mathrm{H}_{{m}} (z)\,,
\end{align}
where $m$ is a multi-index,  ${m}=\{m_1,\ldots, m_p\}$ with the entries $m_k\in\{0,1\}$. 
The power $a$ takes integer values, $a=0,1$ for quarks and $a=0,1,2$ for gluons.
Gegenbauer moments of the one-loop CFs have been calculated in 
Refs.~\cite{Melic:2002ij,Kumericki:2007sa}  by first computing  the Mellin moments 
and then performing a resummation. 
We will describe a different approach and use it for the two-loop CFs.
This possibility was also mentioned in~\cite{Melic:2002ij}.

\subsection{General procedure }\label{subsect:general}%
For definiteness, we consider quark distributions.
The starting observation is that the polynomials 
\begin{align}
G^{(3/2)}_N(z)= 2 z\bar z C_{N-1}^{(3/2)}(1-2z)
\end{align}
are eigenfunctions of $\mathrm{SL}(2,R)$ invariant
operators $\mathbb  H$:%
\footnote{By definition, $\mathrm{SL}(2,R)$-invariant operators commute with the generators of collinear conformal transformations which
depend on the conformal spin: $j=1$ for quarks and $j=3/2$ for gluons.}
\begin{align}\label{HS}
\int_0^1 \!dz'\, \mathbb H(z,z')\, G^{(3/2)}_N(z') =E_N\, G^{(3/2)}_N(z)\,.
\end{align}
Thus if $M_N[f]$ is the Gegenbauer moment of a function $f(z)$,
\begin{align}\label{MNf}
M_N[f]=\int_0^1 dz f(z)\,G^{(3/2)}_N(z),
\end{align}
then the Gegenbauer moment of the function
\begin{align}\label{fHH}
f^{\mathbb H}(z)=\int_0^1 dz' f(z')\mathbb H(z',z)
\end{align}
is  given by the product of $M_N[f]$ and $E_N$,
\begin{align}
M_N[f^{\mathbb H}]=M_N[f] E_N\,.
\label{FirstTry}
\end{align} 
This equation can be used to construct a basis of functions $f^{\mathbb H}(z)$ for which the Gegenbauer moments are known
and the DVCS CFs can then be written as linear combinations of these functions. 
The problem is that explicit expressions for the invariant operators $\mathbb H(z',z)$ in momentum fraction representation 
can be rather involved. Going over to the position space proves to be crucial to make this technique feasible in practice.

The action of the $\mathrm{SL}(2,R)$ invariant operator on a function of two variables 
$\varphi(w_1,w_2)$ in position space can be written as (see, e.g., \cite{Braun:2017cih}) 
\begin{align}
[\mathbb H^{(h)} \varphi](w_1,w_2) &=\int_0^1d\alpha\int_0^{\bar\alpha}d\beta\, h(\tau)\varphi(w_{12}^\alpha,w_{21}^\beta)\,,
\label{InvOperatorQuark}
\end{align}
where 
\begin{align}
w_{12}^\alpha & = w_1\bar\alpha+w_2\alpha\,, \qquad \bar\alpha = 1-\alpha\,, 
\end{align} 
and $\tau =\alpha\beta/(\bar\alpha\bar\beta)$ is the so-called conformal ratio. 
Eigenfunctions of such operators, 
\begin{align}
 [\mathbb H^{(h)} \Psi^{p}_N](w_1,w_2) = E_N^{(h)}  \Psi^{p}_N(w_1,w_2)\,, 
\end{align}
can be chosen as
\begin{align}\label{Psipj}
\Psi^{p}_N(w_1,w_2)&=\int_0^1 \!dz\, e^{-ip(w_1 z+w_2\bar z)}\, G^{(3/2)}_N(z)
\end{align}
for an arbitrary $h(\tau)$.
The corresponding eigenvalues are given by
\begin{align}\label{Eigenvalues}
E_N^{(h)}&=\int_0^1d\alpha\int_0^{\bar\alpha} d\beta\, h(\tau)\, (1-\alpha-\beta)^{N-1}\,.
\end{align}
Since the eigenvalues do not depend on $p$,
integrating Eq.~\eqref{Psipj} over $p$ with a certain weight function 
one obtains the eigenfunctions of the form
\begin{align}\label{psif}
\Psi^{(f)}_N(w_1,w_2) &= \int_0^1 \!dz\, f(w_1 z+w_2 \bar z)\, G^{(3/2)}_N(z)\,.
\end{align}
where the function $f(z)$ is more or less arbitrary.
Applying the operator $\mathbb H^{(h)}$ to \eqref{psif} and sending $w_1\to 1$ and $w_2\to 0$ at the end,
one immediately finds that the Gegenbauer moments of the functions $f(z)$ and
\begin{align}\label{ftof}
f^{h}(z)&= [\mathbb H^{(h)}f](z) =
\int_0^1d\alpha\int_0^{\bar\alpha} d\beta\, h(\tau)\, f( z\bar\alpha + \bar z \beta)
\end{align}
are related to each  other as follows
\begin{align}
M_N[f^{h}]=E^{(h)}_N M_N[f]\,,
\label{master}
\end{align}
cf. \eqref{FirstTry}.
Thus the mapping~\eqref{ftof}, $\mathrm T_{h}: f\mapsto f^{h}$, allows one to construct functions with  known
Gegenbauer moments using position-space kernels $h(\tau)$. In this way construction of the 
(much more complicated) invariant operators in momentum fraction space is avoided. 
Note that any two mappings commute:
\begin{align}
\mathrm T_{h_1}\mathrm  T_{h_2} =\mathrm T_{h_2} \mathrm T_{h_1}=\mathrm T_{h_1\otimes h_2}\,,
\end{align}
so that this construction can be applied recursively. 
It turns out to be possible to generate all necessary functions~\eqref{HPLa} starting from $f(z)=1/z$
and using sufficiently simple kernels $h(\tau)$.

\subsection{Quark CFs }\label{subsect:QCFs}
Here we explain in detail how this approach works for the case of quark CFs. 
Our aim is to find the Gegenbauer moments \eqref{MNf} of the functions
\begin{align}\label{FbarF}
F^\pm_{{m}}(z)=\frac1z \mathrm H_{{m}}(z) \pm \frac1{\bar z} \mathrm H_{{m}}(\bar z)\,,
\notag\\
\bar  F^\pm _{{m}}(z)=\frac1{\bar z } \mathrm H_{{m}}(z) \pm \frac1{ z} \mathrm H_{{m}}(\bar z)\,,
\end{align}
where ${m}=\{m_1,\ldots, m_t\}$, with $m_k$ taking two values, zero and one.
Since
\begin{align}
\mathrm H_{m}(\bar z)=
(-1)^t \mathrm H_{1-m}(z)+\text{lower weight functions}
\end{align}
one can always assume that the right-most index is zero, $m_t=0$. 
For uniformity of notation we define $\mathrm H_{\{\}}$ for the empty set $m=\{\}$ as 
\begin{align}
\mathrm H_{\{\}}&= 1\,.
\end{align}
Let also
\begin{align}
  M^\pm_{{m}}(N) \equiv M_N[F^\pm_{{m}}]\,,
\qquad 
 \widebar M^\pm_{{m}}(N) \equiv  M_N[\widebar F^\pm_{{m}}]\,.
\end{align}
The moments $M^\pm_{{m}}(N)$, $\widebar{ M}^\pm_{{m}}(N)$ are non-zero for odd (even) $N$, respectively.
For the empty set obviously  
\begin{align}
 M^\pm_{\{\}}(N) &= \phantom{\pm} 1\,, \qquad N = 2,4,\ldots
\notag\\
{\widebar M}^\pm_{\{\}}(N) &=\pm 1\,, \qquad N=1,3,\ldots
\end{align}
cf. \eqref{LO}. 

As the next step, let us calculate Gegenbauer moments for the functions of weight one, $m = \{0\}$.
To this end, first apply the operator $\mathbb H^{(1)}$, i.e. with the kernel $h(\tau) =1$ to the
function $F^\pm_{\{\}}(z)= 1/z\pm 1/\bar z$. One easily finds
\begin{align}
\mathbb H^{(1)} F^\pm_{\{\}}(z) = -\left(\frac{\ln z}{\bar z}\pm \frac{\ln\bar z}{z}\right)= - {\bar F}^\pm_{\{0\}}(z).
\end{align}  
The eigenvalues of $\mathbb H^{(1)}$ \eqref{Eigenvalues} are $E_N^{(1)} = 1/N(N+1)$. Thus, using Eq.~\eqref{master},
one gets
\begin{align}
{\widebar M}^\pm_{\{0\}}(N) = - \frac{1}{N(N+1)}.
\label{first}
\end{align}
A special invariant operator with the eigenvalues given by $2S_1(N)$ takes the form
\cite{Braun:2009vc}
\begin{align}
[\widehat{\mathbb H} f] (w_1,w_2)  &=  \int_0^1 \frac{d\alpha}{\alpha}\Big[2f(w_1, w_2)
\notag\\
&\quad -
\bar\alpha\big( f(w_{12}^\alpha,w_2) + f(w_1,w_{21}^\alpha)\big) \Big]\,.
\label{Hhat}
\end{align} 
It can be written in the form \eqref{InvOperatorQuark} with the kernel $h(\tau) = \delta_+(\tau)$ \cite{Braun:2017cih} 
(up to a constant), but the above explicit expression is more convenient for applications.
One easily obtains 
\begin{align}
[\widehat{\mathbb H}F^\pm_{\{\}}](z) =-\frac{\ln z\pm \ln \bar z}{z\bar z}
= - {F}^\pm_{\{0\}}(z) -{\bar F}^\pm_{\{0\}}(z)
\end{align}
that results in the following relation for the moments,
see Eq.~ \eqref{master}
\begin{align}
 {M}^\pm_{\{0\}}(N) + {\widebar M}^\pm_{\{0\}}(N) = - 2S_1(N)\,.
\label{second}
\end{align} 
Taking into account~\eqref{first} one derives 
\begin{align}
 {M}^\pm_{\{0\}}(N) = - 2S_1(N) + \frac{1}{N(N+1)}\,.
\end{align} 
Thus we have obtained moments of the weight-one functions \eqref{FbarF} 
avoiding a direct calculation of the integral~\eqref{MNf}.

In order to proceed to the next level we apply the 
operators $\mathbb H^{(1)}$ and $\widehat {\mathbb H}$ to the functions of the weight one, which Gegenbauer moments we 
know already. These give us three new functions of weight two with known moments, not four, because 
the operators  $\mathbb H^{(1)}$ and $\widehat {\mathbb H}$ commute. Thus we end up with three equations for the 
four existing weight-two functions, and need to add one more operator to our basis to compensate for this mismatch.
A large list of invariant operators and the corresponding eigenvalues up to weight five can be found in~\cite{Ji:2023eni}. 
We need to choose one of them with the eigenvalue given by harmonic sums of weight two. 
One finds by inspection  that $\mathbb H^{(\bar \tau)}$, i.e. the invariant operator with 
the kernel $h(\tau) = \bar\tau$, has the required complexity, with the eigenvalues $(-1)^N[2S_{-2}(N)+\zeta_2]$. 
Calculating the integral~\eqref{ftof} with $f=F^{\pm}_{\{\}}$  and the kernel $h=\bar \tau$ we obtain the
necessary fourth equation. 

This procedure can be continued iteratively to the functions of higher weight.
At weight four, eight independent harmonic polylogarithms 
exist if the last index is fixed to be zero, and we need two combinations as in Eq.~\eqref{FbarF}.
Thus we need to construct sixteen 
independent basis functions, alias sixteen invariant operators using which such functions can be generated. 
We collect these operators and their eigenvalues in the Appendix. 
  
The feasibility of this technique relies heavily on the fact  that the integrals~\eqref{ftof}
with functions $f$ defined in Eq.~\eqref{FbarF} and the kernels
$h(\tau)$ of the form
\begin{align}
h(\tau)=\left\{\bar\tau\,
\mathrm H_{m}(\tau),\, ({\bar\tau}/{\tau}) \mathrm H_{m}(\tau)\right\}
\end{align}
can be calculated easily with the \texttt{HyperInt} package~\cite{Panzer:2014caa}. The result of the integration
is given in a general case by a linear combination
of the functions defined in Eq.~\eqref{ftof},
\begin{align}
I_h[f]=\sum_m (C_m^h F_m^{\pm} + \bar C_m^h \bar F_m^{\pm}).
\end{align}
The Gegenbauer moment of this combination  is given by the product of the Gegenbauer moment of the function $f$ 
and the eigenvalue of the kernel $\mathbb H^{(h)}$.
The eigenvalues can be calculated using recurrence relations in $N$, up to the  weight  five  they can be 
found in~\cite{Ji:2023eni}. 
In this way one can generate  a sufficient number of equations 
in order to determine  the Gegenbauer moments $M^\pm_{{m}}(N),\widebar{ M}^\pm_{{m}}(N)$ of the functions in Eq.~\eqref{FbarF},
up to weight four (and, in principle, also higher).

In addition to the functions defined in \eqref{FbarF}, the two-loop DVCS CFs contain harmonic polylogarithms 
$\mathrm H_{m}$ without the $z$ or $\bar z$ factor in the denominator.  
Such contributions can be handled  using a recurrence relation for the Gegenbauer polynomials that results in
\begin{flalign}
M_{N-1}\big[(\bar z\!-\!z)f\big]&= \frac{N M_{N}\big[f\big]\!+\!(N+1)M_{N-2}\big[f\big]}{2N+1}.
\end{flalign}

\subsection{Gluon CFs }\label{subsect:GCFs}

Extension of the technique described above to the gluon CFs is straightforward. To this end we need to calculate 
moments 
\begin{align}
M_N[f] &= \int_0^1\! dz\, f(z) G^{(5/2)}_N(z)\,.
\end{align}
of the functions defined in Eq.~\eqref{FbarF} and, in addition, 
\begin{align}\label{GbarG}
G^\pm_{{m}}(z)=\frac1{z^2} \mathrm H_{{m}}(z) \pm \frac1{\bar z^2} \mathrm H_{{m}}(\bar z)\,,
\notag\\
\bar  G^\pm _{{m}}(z)=\frac1{\bar z^2 } \mathrm H_{{m}}( z) \pm \frac1{ z^2} \mathrm H_{{m}}( \bar z)\,,
\end{align}
with respect to the polynomials 
\begin{align}\label{Rmoments}
G^{(5/2}_N(z)& = 6(z \bar z)^2 C_{N-2}^{(5/2)}(1-2z)\,.
\end{align}
In this case invariant kernels can be written in the form $(1-\alpha-\beta)h(\tau)$,%
\footnote{For the general case of a two-particle operator with conformal spins $j_1$ and $j_2$,
the invariant kernels have the form $\bar\alpha^{2j_1-2}\bar\beta^{2j_2-2}h(\tau)$, see, e.g., 
in \cite[Eq.~(2.40)]{Braun:2009vc}.
}
 so that invariant operators and the corresponding eigenfunctions are written as 
\begin{align}
[\mathbb H \varphi](w_1,w_2) &=\int_0^1d\alpha\!\int_0^{\bar\alpha}\!d\beta\, 
(1\!-\!\alpha\!-\!\beta)\,h(\tau)\varphi(w_{12}^\alpha,w_{21}^\beta)\,,
\notag\\ \Psi^{p}_N(w_1,w_2)&=\int_0^1 dz e^{-ip(w_1 z+w_2\bar z)}  G^{(5/2)}_N(z)\,,
\label{InvOperatorGluon}
\end{align}
cf. \eqref{InvOperatorQuark}.
The $\widehat{\mathbb{H}}$ operator for gluons has the same form as for quarks \eqref{Hhat}, 
with $\bar\alpha$ factors replaced by $\bar\alpha^2$, see \cite[ Eq.~(4.1)]{Braun:2009vc}. 
The eigenvalues are given by the same expression as for quarks, Eq.~\eqref{Eigenvalues}, 
but the relation~\eqref{ftof} acquires an extra $(1-\alpha-\beta)$ factor as well,
\begin{align}\label{ftofgluon}
f^{h}(z)&=\int_0^1d\alpha\!\int_0^{\bar\alpha}\! d\beta\,(1-\alpha-\beta)\, h(\tau) f(w_{12}^\alpha z + w_{21}^\beta \bar z)\,.
\end{align}
With these modifications, the relation between the Gegenbauer moments of $f^{h}(z)$ and $f(z)$ remains the same, 
Eq.~\eqref{master}. 

In our normalization, the Gegenbauer moments of the lowest weight functions are
\begin{align}
M_{N}\left[F^{\pm}_{\{\}}\right] &=1, 
\notag\\
M_{N}\left[G^{\pm}_{\{\}}\right] &=(N-1)(N+2)\,.
\end{align}
Note that $"+"$ moments are nonzero for even $N$, and $"-"$ moments for odd  $N$. 
Proceeding exactly as in the quark case we have calculated 
the Gegenbauer moments~\eqref{Rmoments} of the functions $F(\bar F)^\pm_{m}, \, G(\bar G)^\pm_{m}$  up to weight four.

The final results for the CFs are collected in the next section and also in Mathematica format
as supplementary material. \cite{Supplementary}

\section{Results}\label{sect:results}
In this section we present explicit expressions for the Gegenbauer moments \eqref{GegMoments} of the DVCS 
CFs~\eqref{Hfactorform},\eqref{CqCg}.
The quark CFs can conveniently be separated into non-singlet and singlet contributions with a different dependence 
on electric charges 
\begin{align}
C_q &= e_q^2\, C_{q,\NS}  + \Sigma_Q\, C_{q,\PS}\,,
\notag\\
\widetilde C_q &= e_q^2\, \widetilde C_{q,\NS}  + 
\Sigma_Q \, \widetilde C_{q,\PS}\,,
\end{align}
where $\Sigma_Q=\frac12 \sum_{q'} e_{q'}^2$.
They can further be separated in the contributions of different orders of perturbation theory and color structures,
\begin{align}
C_{q,\NS} &= C_{q,\NS}^{(0)} + a_s C_F C_{q,\NS}^{(1)}
\notag\\
&
+ a_s^2 C_F\Big(  C_F C_{q,\NS}^{(2)\CF}
+ \frac1{N_c} C_{q,\NS}^{(2)\NP}
+  \beta_0 C_{q,\NS}^{(2)\beta} \Big) +\ldots\,,
\notag\\
C_{q,\PS} & =a_s^2 C_F C_{q,\PS}^{(2)} +\ldots\,,
\end{align}
where $C_F = (N_c^2-1)/(2N_c)$, $a_s = \alpha_s/(4\pi)$ and $\beta_0 = 11/3 N_c - 2/3 n_f$. 

The gluon CFs start at one loop and can be decomposed in different contributions as
\begin{align}
C_g &=\Sigma_Q\Big\{
a_s C_g^{(1)}\! +\!a_s^2\Big(C_F C_ {g}^{(2)\CF}\! +C_A C_{g}^{(2)\CA}\Big)+\ldots
\Big\}.
\end{align}
The decomposition of the axial-vector CFs (with a ``tilde'') is the same.

The expressions presented below for Gegenbauer moments \eqref{GegMoments} 
of the vector CFs  are for even $N$, and for the axial-vector CFs for odd $N$, respectively.
We use the notation $L=\ln Q^2/\mu^2$ and
\begin{align}
H=\frac{1}{N(N+1)}\,.
\end{align}
All harmonic sums are assumed to have the argument $N$, e.g., $S_1 \equiv S_1(N)$.   

The tree-level expressions in the chosen normalization  \eqref{GegMoments} are
\begin{align}
C_{q,\NS}^{(0)}(N) = 1\,, \qquad  \widetilde{C}_{q,\NS}^{(0)}(N) = 1\,.
\end{align}
At one loop one obtains \cite{Mueller:2005nz,Kumericki:2007sa}
\begin{align}
C_{q,\NS}^{(1)} &=4 S_1^2-4 S_1 H\! +\!2 H^2 +5 H\! -\!9\! - 
 L \Big(4S_1-2H- 3\Big),
\notag\\
\widetilde C_{q,\NS}^{(1)} &=4 S_1^2-4 S_1 H\! +\!2 H^2 +3 H\! -\!9\! - 
L \left(4S_1-2H- 3\right),
\notag\\
C_g^{(1)} &=		-( 2 H + 1) \Big (S_1 -\frac12 H \Big) 
			- \frac12 
			+ L \Big( \frac12 + H \Big) \,,
\notag\\
\widetilde C_g^{(1)} &=  + ( 2 H-1 ) \Big( S_1 - \frac12 H + \frac12 \Big) 
			 +  L \Big( \frac12 - H \Big) \,.
\end{align}
The coefficient of the logarithm for the quark distributions is of course 
related to the one-loop anomalous dimension
\begin{align}
\gamma_{qq}^{(1)}(N)  =2 C_F \left(4S_1-2H -3\right)\,.
\end{align} 
The two-loop moments are considerably more complicated. We obtain, for quarks
\begin{widetext}
\begin{align}
C_{q,\NS}^{(2)\beta}&=	\frac43 S_3 
		      + \frac83 S_1^3 
		      + \frac{20}3 S_1^2 
		      + 4 S_{-2} 
		      + \frac{38}9 S_1
		      - \zeta_3
		      - \frac83\zeta_2
		      - 4 S_1^2 H 
		      + 2 S_{-2} H
		      + \frac{10}3 S_1 H 
		      + 8 S_1 H^2 
		      + \zeta_2 H
                      - 5 H^3
\notag\\
&\quad
		       - 5H^2 
		      + \frac{121}{18} H 
		      - \frac{457}{24}  
		      + L\left( -4S_1^2
		      + 4 S_1 H
		      - \frac{20}3 S_1 
		      - 4H^2
		      - \frac53 H 
		      + 2\zeta_2
		      + \frac{19}2\right) 
		      + \frac12 L^2  \Big(4S_1-2H -3\Big)\,,
\notag\\[2mm]
{\widetilde C}_{q,\NS}^{(2)\beta} 
                      &=\frac43 S_3 
		      + \frac83 S_1^3 
		      + \frac{20}3 S_1^2  
		      + \frac{38}9 S_1
		      - \zeta_3
		      - \frac{14}3\zeta_2
		      - 4 S_1^2 H  
		      - 2 S_{-2} H 
		      - \frac{2}3 S_1 H 
		      +  8 S_1 H^2 
		      - \zeta_2 H 
		      - 5 H^3 
		      - H^2
\notag\\
&\quad
		      + \frac{115}{18} H 
		      - \frac{457}{24}   
		      + L\left(
		      - 4 S_1^2
		      + 4 S_1 H 
		      - \frac{20}3 S_1 
		      -4 H^2 
		      + \frac13 H 
		      + 2\zeta_2 
		      +\frac{19}2
			\right)  
		      +  \frac12 L^2  \Big(4S_1-2H -3\Big)\,,
\end{align}
\begin{align}
C_{q,\NS}^{(2)\CF}&= 	     8 S_1^4 
		    -8 \left(2 S_{1, 3}-S_4\right) 
		    + 32 S_3 S_1 
		    - 4 S_{-2}^2  
		    + 8 (2 S_{-2, 1}-S_{-3})
		    - 8 H\left( S_3 
		    + S_{-2} S_1 
		    + 2 S_1^3 
		    - 4 S_1^2
		     \right)
\notag\\&\quad
		    + 4 H^2\left (2 S_{-2} 
		    + 7 S_1^2 \right )
		    -\frac{76}3 S_1^2 
		    - 12 S_{-2} 
		    - S_1\left(32 H^3 
		    + 92  H^2  
		    + \frac{86}3  H 
		    - \frac{128}9 \right)
 		    - 4\zeta_2 ( S_{-2}
		    + 4 S_1^2
	            - 3 S_1 H 
	            + 3 S_1) 
\notag\\
&\quad
	            - 64\zeta_3 S_1
		    - 2\zeta_2 H(H+3) 
		    + 32\zeta_3 H
		    + 10 H^4 
		    + 50 H^3  
		    + 49 H^2
	            - \frac{815}{18} H
	            - \frac{14}5\zeta_2^2 
	            + 43\zeta_3
	            + \frac{701}{24} 
	            + \frac{16}3\zeta_2
\notag\\
&
\quad
		    + L\biggl(
		    - 16 S_1^3
		    - 8 S_3 
		    + 12 S_1^2 (2H+1) 
		    + 8 H S_{-2}
		    + 4 S_1\left(\frac{19}3 
		    - 10 H 
		    - 6 H^2 
		    + 4 \zeta_2
		    \right)
		    + 8 H^3
		    + 32 H^2 
		    + \frac{67}3 H 
		    - \frac{47}2
\notag\\
&
\quad
		    - 4\zeta_2 H 
		    - 6\zeta_2  
		    + 4\zeta_3
			\biggr)
		    + \frac12 L^2  \Big(4S_1-2H -3\Big)^2\,,
\notag\\[2mm]
{\widetilde C}_{q,\NS}^{(2)\CF}&=  8 S_1^4 
		      - 8 (2 S_{1, 3}-S_4) 
		      + 32 S_3 S_1 
		      - 4 S_{-2}^2  
		      - 8 H \left( S_3 
		      - S_{-2} S_1
		      + 2 S_1^3
		      - 3 S_1^2\right)
		      + 4 H^2 \left(
		      - 2 S_{-2} 
		      + 7 S_1^2
		      \right)
\notag\\
&\quad
		      - \frac{76}3 S_1^2 
		      - S_1\left( 32  H^3 
		      + 76  H^2   
		      + \frac{26}3  H 
		      - \frac{128}9 \right)
		      - 4\zeta_2 ( S_{-2}
		      + 4 S_1^2
		      - 5 S_1 H  
		      + 3 S_1) 
		      - 64\zeta_3 S_1
\notag\\
&\quad
		      - 2 \zeta_2 H(5H+1) 
		      + 32\zeta_3 H 
		      + 10 H^4 
		      + 42 H^3 
		      + 21 H^2
		      - \frac{1001}{18}  H 
		      + \frac{701}{24} 
		      - \frac{14}5\zeta_2^2 
		      + 39\zeta_3 
		      + \frac{34}3\zeta_2
\notag\\
&\quad
			+L\biggl(
			- 16 S_1^3 
			- 8 S_3 
			+ 12 S_1^2 (1+2H) 
			+ \frac{76}3 S_1 
			+ 16\zeta_2 S_1 
			+ 4\zeta_3 
			- 6 \zeta_2
			- \frac{47}2
			+ H \left( -32 S_1
			- 8 S_{-2} 
			- 12 \zeta_2
			+ \frac{49}3
			\right)
\notag\\
&\quad
			+ H^2 \big( 28 
			- 24 S_1 \big)
			+ 8 H^3
			\biggr)
			+  \frac12 L^2  \Big( 4 S_1 -2 H -3 \Big)^2\,,
\end{align}
and
\begin{align}
C_{q,\NS}^{(2)\NP}& =\frac{6 \left(4S_{-2}+3\right)}{(N-2)(N+3)} 
		- 4(2S_{1,3}- S_4) 
		- 4S_{-4} 
		- 8 S_{-2}^2 
		+ 16(2S_{1,-2}-S_{-3}) S_1
		+ 24 S_1 S_3 
		+ 4\zeta_2 S_{-2}
\notag\\
&\quad
		+ 8\zeta_2 S_1^2 
		- 52\zeta_3 S_1 
		+ 8 S_{-2} 
		+ \frac{16}3 S_1^2 
		+ \frac{64}9 S_1
		+ \frac{2}{5}\zeta_2^2
		+ 54\zeta_3
		- \frac {10}3\zeta_2
		- \frac{73}{12} 
		+ H^2\Big(32 S_{-2} +10\Big)
\notag\\
&\quad 		+H\left( 8(2 S_{-2,1}-S_{-3})
		- 40 S_{-2}S_1
		- 12 S_3
		+ 36\zeta_3
		- 8\zeta_2 S_1
		+ 52 S_{-2}
		- \frac{76}3 S_1 
		+ \frac{1}9
\right)
\notag\\
&\quad 		+ L\left( 
		- 8(2 S_{1,-2}-S_{-3}) 
		- 8 S_3 
		- \frac{16}3 S_1 
		+ 8 H S_{-2} 
		+ \frac{20}3 H
		+ 1
		\right),
\notag\\[2mm]
{\widetilde C}_{q,\NS}^{(2)\NP}&=\frac{24 \left(S_{-2}+ 1\right)}{(N-1)(N+2)}
 		+ 2 \biggl\{12 S_1 S_3 
          	- 2 S_{-4} 
           	- 2 ( 2 S_{1,3}-S_4 )
         	- 4 S_{-2}^2 
          	+ 8 ( 2 S_{1,-2}-S_{-3} ) S_1
          	+ 2 ( \zeta_2 + 2 ) S_{-2} 
	  	+ 4 \zeta_2 S_1^2
\notag\\
&\quad
		+ \frac83 S_1^2 
		- 26 \zeta_3 S_1 
		+ \frac{32}9 S_1 
		+ \frac15\zeta_2^2 
		+ 27\zeta_3
		- \frac53\zeta_2 
		- \frac{73}{24}
		+ 6 H^4
		+ 12 H^3
		- 8 S_1 H^3
		+ 2 H^2 \Big( 2S_{-2} 
			- 8 S_1 
			+ \zeta_2 
			- 5\Big)
\notag\\
&\quad
 		+ H\left( -2 S_{3} 
 		  	  + 4 ( 2 S_{-2,1}-S_{-3} ) 
 		  	  - 12 S_{-2} S_1 
 		  	  - 2 S_{-2} 
 		  	  - 4 \zeta_2 S_1 
 		  	  - \frac{ 2 }{ 3 } S_1
			  + 8 \zeta_3 
			  + 2 \zeta_2
			  - \frac{113}{18}
\right)
\biggr\}
\notag\\
&\quad 
		+ L \Big(- 8 ( 2 S_{1,-2} - S_{-3} ) 
			 - 8 S_3 
			 - \frac{16}3 S_1 
			 + 8 H S_{-2} 
			 + 8 H^3 
			 + 16 H^2 
			 +\frac{20}3 H
			 +1 \Big).
\end{align}
For the pure-singlet quark CFs we obtain:
\begin{align}
C_{q,\PS}^{(2)} & = 	\frac{32}{(N-1)(N+2)}\left( 
                 	  2 S_1^2 
                 	- S_{-2} 
                 	- 4 S_1 
                 	- \zeta_2
                 	+ 3\right) 
                 	+ H \Big(
			- 32 S_1 S_{-2} 
			- 16 ( 2 S_{-2,1} - S_{-3} ) 
			- 48 S_1^2 
			- 24 S_{-2}
\notag\\
&\quad
			+ 144 S_1
			- 16 \zeta_2 S_1
			- 8 \zeta_3 
			+ 4 \zeta_2 
			- 84 
			- H \left( 32 S_1^2 
			- 16 S_{-2}
			- 160 S_1 
			- 8 \zeta_2 
			+ 140\right)
			+ 4 H^2 (16 S_1 - 33 )
			- 40 H^3
\Big)
\notag\\
&\quad
			+ 16 L\,\left(
			- \frac{4 (S_1-1)}{(N-1)(N+2)} 
			+ H\left(  2 S_{-2}  
			+ \Big( 2 H 
			+ 3  \Big) S_1 
			+ \zeta_2  
			- 2 H^2 
			- 5 H
			- \frac92  
			\right) 
			\right)
\notag\\
&\quad
			+ 4 L^2 \left( \frac{4}{(N-1)(N+2)} 
			- 2 H^2
			- 3 H\right)\,,
\notag\\[2mm]
{\widetilde C}_{q,\PS}^{(2)} & =	 \frac{ 32 ( S_{-2} + 1 ) }{ (N-1) (N+2) } 
			- 4 H\Big( 8 S_1 S_{-2} 
			+ 4 ( 2 S_{-2,1} - S_{-3} ) 
			+ 4 S_1^2 
			+ 18 S_{-2} 
			+ 8 S_1 
			+ 4 \zeta_2 S_1 
			+ 2 \zeta_3 
			+ 5 \zeta_2 
			+ 14
\Big)
\notag\\
&\quad
			+ 8 H^2 \left( 4 S_1^2 
			+ 2 S_{-2} 
			- 4 S_1 
			+ \zeta_2 
			- 8 \right)  
			+ 4 H^3 (3-16 S_1) 
			+ 40 H^4
\notag\\
&\quad
			+ 16 L\,H\, \Big( 2 S_{-2}  
			+ S_1 (1-2H) 
			+ \zeta_2 
			+ 1  
			+ H 
			+ 2 H^2 \Big)
			+ 4 L^2 H\left( 2H - 1\right).
\end{align}
Gegenbauer moments of the two-loop gluon CFs are given by the following expressions:
\begin{align}
C_{g}^{(2),\CF} &= 	- L^2  (2H+1) \left( S_1 
			- \frac{1}{4} ( 3 + 2 H ) \right) 
			+ L \biggl\{ ( 2 H + 1) \bigg( 4 S_1^2
			- \zeta_2
			+ 2 H^2
			+ 4 H
			- 2 \bigg) 
			- 8 H (H+1) S_1 
			+ 6 H \biggr\}
\notag\\
&\quad
			- \frac{ 2 ( 4 S_{-2} + 3)}{ (N-2)(N+3) }
			+ ( 1 - 8 H ) \Big( 2 S_{-2,1} - S_{-3}\Big) 
			+ 16 S_1 S_{-2} 
			- \frac53 (1-4 H) S_3 
			- \frac{10}3 S_1^3 
			+ \Big( 10 H^2 + 5 H - 3 \Big) S_1^2
\notag\\
&\quad
			- H ( 16 + 21 H) S_{-2} 
			+ \zeta_2 ( 2 H + 1) S_1 
			+ \left( \frac52 
			- 3 H 
			- 24 H^2 
			- 12 H^3\right) S_1 
			+ \big( 7 - 10 H\big) \zeta_3 
			+ \zeta_2 \left( \frac14 - H - H^2 \right)
\notag\\
&\quad
			+ \frac14 \Big (-2
			-52 H
			+ 49 H^2
			+ 68 H^3
			+ 20 H^4
			\Big)\,,
\notag\\[2mm]
\widetilde C_{g}^{(2)\CF} &= L^2 (2H-1) \left( S_1 
			- \frac{1}{4}( 3 + 2 H )
			\right)
			- L \bigg\{ (2H-1) \Big( 4 S_1^2 
			- 4 H S_1 
			+ 2 H^2
			+ H
			-\zeta_2
			-2 \Big)
			\bigg\}
\notag\\ &\quad
			+ \frac{ 24 ( S_{-2} + 1 ) }{ (N-1)(N+2) }
			+ ( 1 + 8 H ) \Big( 2 S_{-2,1} - S_{-3} \Big)
			- \frac{5}{3} ( 1 + 4 H ) S_3
			- 16 H S_{-2} S_1 
			+ 2 ( 3 H - 1) S_{-2}
\notag\\&\quad
			+ ( 2 H - 1 )\left( \frac{10}3 S_1 
			- 5 H 
			+ 3 \right) S_1^2
			+ \left( \frac52 
			- 9 H 
			+ 12 H^3  
			+ ( 1 - 2 H )\zeta_2
			\right) S_1 
			+ \zeta_2\left( H^2
			+ H 
			- \frac34
			\right)
\notag\\
&\quad
			+ \zeta_3 ( 10 H + 7 ) 
			+ \frac14 \Big( - 2
			- 100 H
			+ 33 H^2
			+ 8 H^3
			- 20 H^4
			  \Big)\,,
\end{align}
and
\begin{align}
C_{g}^{(2)\CA} &= 	L^2 \left(
			 \frac6{(N-1)(N+2)} 
			 - ( 2 H + 1 ) S_1 
			 - 2 H ( H +2 )\right) 
			 - 2 L\biggl\{
			   (N-1)(N+2)\left(  2 S_{-2,1}
			   - S_{-3}
			   + \frac12\zeta_3
			\right)
\notag\\
&\quad
			- ( 2 H + 1 ) \Big( S_1^2 
			+ \zeta_2 \Big) 
			- ( 1 + 4 H ) S_{-2} 
			+ \frac{ 12 ( S_1 - 1 )}{ (N-1)(N+2) }
			- \Big( 2 + 7 H + 2 H^2 \Big) S_1
			+ 4 H^2 ( H + 3 )
			+ \frac{25}2 H 
			- \frac12
\biggr\}
\notag\\
&\quad
			+ \frac{ 12 ( 2 S_1^2 - S_{-2}
			- \zeta_2
			- 4 S_1 +3 )}{ (N-1)(N+2) } 
			+ 6 (N\!-\!1)(N\!+\!2)\biggl\{ 2 S_{-2,1,1}
			- S_{-3,1}
			- S_{-2,2}
			+ \frac23\Big( 2 S_{1,-2,1}
			- S_{-3,1}
			- S_{1,-3} \Big)
\notag\\
&\quad 
			+ \frac12 S_{-4} 
			+ \frac13 \left( S_{-2,1} 
			- \frac12 S_{-3} \right)
			+ \frac13 \zeta_2 S_{-2}
			+ \frac{2}{15} \zeta_2^2 
			+ \frac13 \zeta_3 S_1
			+ \frac1{12} \zeta_3
			\biggr\}
 			- \Big( 1 + 2 H \Big) S_1^3  
 			- \frac13 \Big( 1 + 14 H \Big) S_3
 \notag\\
 &\quad
			- 8 H S_{-2} S_1
			- 6 \left( S_{-2,1} 
			- \frac12 S_{-3} \right) 
			- \Big( 4 H^2
			+ 14 H 
			+ 5 \Big) S_1^2 
			+ 2 \left( H^2
			+ H
			- \frac54
			\right)\zeta_2
			+ 2 \Big( 3 H - 2 \Big) \zeta_3 
			+ 4 \zeta_2 H S_1
 \notag\\
 &\quad
			+ \left( 8 H^2 
			- 2 H 
			- \frac12 \right) S_{-2} 
			+ S_1 \left( 12 H^3 
			+ 42 H^2
			+ 46 H 
			- \frac{11}2 \right) 
			- 10 H^4
			- 38 H^3
			- \frac{95}2 H^2
			- 29 H - 2\,,
\notag\\[2mm]
\widetilde C_{g}^{(2)\CA}&=L^2 (2H-1) \biggl( S_1 - 2 H \biggr)
			- 2 L \biggl\{ (N-1)(N+2) \biggl( 2 S_{-2,1} 
			- S_{-3}
			+ \frac{1}{2} \zeta_3 \biggr)
			+  ( 2 H - 1 ) \biggl( S_1^2 
			+ \zeta_2 \biggr)
\notag\\
&\quad 
			+ \left( 4 H - 1 \right) S_{-2} 
			+ H ( 3 - 10 H ) S_1 
			- \frac12 
			+ \frac52 H 
			+ 2 H^2 
			+ 6 H^3
\biggr\}
			+ 2 (N-1 )(N+2) \biggl\{
			- \Big( 2 S_{-2,1} 
			- S_{-3} \Big) S_1
\notag\\
&\quad
			+ \Big(2 S_{1,1,-2}
			- S_{2,-2}
			- S_{1,-3}
			+ \frac12 S_{-4}
				\Big)
			- S_{-2} S_1^2
			+ \frac12 \left( S_{-2,1} 
			- \frac12 S_{-3} \right)
			+ \zeta_2 S_{-2}
			- \zeta_3 S_1
			+ \frac25 \zeta_2^2 
			+ \frac{\zeta_3}{4}
				\biggr\}
\notag\\
&\quad
			- \frac{ 12 ( S_{-2} + 1 ) }{ (N-1)(N+2) }
			+ ( 8 H - 5) \Big( 2 S_{-2,1}
			- S_{-3} \Big) 
			+ 8 H S_{-2} S_1 
			+ \frac13 ( 14 H - 1 ) S_3 
			+ \frac23 ( 2 H - 1 ) S_1^3
\notag\\
&\quad
			+ \left( 22 H 
			- \frac92 \right) S_{-2}
			+ \Big( 2 H
			- 20 H^2
			+ 1 \Big) S_1^2 
			+ 2 S_1 \left(14 H^3
			+ 7 H^2
			+ 5 H
			- \frac74
			- 2 H \zeta_2
			\right)
			- \zeta_3 ( 5 + 2 H) 
\notag\\
&\quad
			+ \zeta_2 \left( 2 H^2
			+ 8 H
			- \frac52 \right)
			- 14 H^4
			- 6 H^3
			+ \frac{41}2 H^2
			+ 17 H
			- 2\,.
\end{align}
\end{widetext}
For completeness, we have calculated also the moments for the two-loop CFs for gluon transversity as defined in Ref.~\cite{Ji:2023xzk}:
\begin{align}
        C_{g,T}(N) &=   
        2\Sigma \biggl\{ a_s
           + a_s^2\Big( 
           - 2 \, L \, C_A \, S_1  + C_F \big(-2 + S_1\big) 
\notag\\&\quad 
   + C_A \big( 1 + 3 S_1 - 2 S_2 + 4 S_{1,1} \big)
            \Big)
\biggr\}.
\end{align}
This expression holds for even $N\geq 2$. 

We stress that the utility of the provided expressions is the 
possibility of analytic continuation to complex $N$ plane.
For fixed integer values of $N$, analytic expressions for the Gegenbauer moments  \eqref{GegMoments}
can be obtained by direct integration using the \texttt{HyperInt} Maple package~\cite{Panzer:2014caa}.
We have checked that the expressions obtained in this way agree identically 
with our results for all CFs for $N\le 20$. 
A numerical comparison using the provided Mathematica expressions was done as well, 
in which case we found agreement to machine precision. 

As the most interesting property, all Gegenbauer moments satisfy the 
reciprocity relation~\cite{Dokshitzer:2005bf,Dokshitzer:2006nm,Beccaria:2009eq}: 
their asymptotic expansion at $N\to \infty$ is symmetric under the interchange  
$N\leftrightarrow -1-N$. This feature is nontrivial and provides additional confidence in the results. It can also be useful in applications. 


\section{Summary and Conclusions }\label{sect:summary}%

We have developed a technique which facilitates calculation of conformal moments of the CFs in DVCS and used it to calculate 
moments of the two-loop CFs. The moments are given by  linear combinations of the reciprocity respecting
harmonic sums~\cite{Dokshitzer:2005bf,Dokshitzer:2006nm,Beccaria:2009eq} and can easily be continued to complex~$N$.
The CFs in this representation are needed for NNLO calculations of the amplitudes using Mellin-Barnes techniques, 
which have several advantages compared to the traditional approach using convolutions in momentum fractions. 
The NNLO contributions in this form can be implemented relatively easily in the existing analysis code~\cite{Gepard} 
and are expected to be substantial~\cite{Braun:2022bpn} for gluon GPD contributions in the EIC kinematic range.  
Since GPDs are normalized to the conventional PDFs in the forward limit, NNLO corrections are also 
necessary for consistency with modern PDF determinations where the NNLO accuracy has become the standard of the field.
 
\vspace{12pt}
\section*{Acknowledgments}
This study was supported by the Deutsche Forschungsgemeinschaft (DFG) Research Unit FOR 2926, ``Next Generation pQCD for
Hadron Structure: Preparing for the EIC'', project number 40824754.



\appendix 
%
%

\section{Invariant kernels }\label{sect:kernels}%

The kernels $h(\tau)$ \eqref{InvOperatorQuark} used in our calculation are collected in Table~\ref{table:1}. 
As already mentioned in the text, a product of two invariant operators with the eigenvalues $E_N^{(h_1)}$ and $E_N^{(h_2)}$ is an invariant operator with the eigenvalues
$E_N^{(h_1\otimes h_2)} = E_N^{(h_1)} E_N^{(h_2)}$. Such products are naturally used to produce operators with the eigenvalues
with higher weights, so that at each step only a few ``new'' operators have to be added.

Note that the same set of functions  $h(\tau)$ can be used for gluon operators. The invariant kernels in this case take the form
$(1-\alpha-\beta)h(\tau)$ but the equation for the eigenvalues, Eq.~\eqref{Eigenvalues}, retains its form.  
The  invariant operator $\widehat{\mathbb{H}}$  has a slightly  different form for quarks and gluons, see \cite[Eq.~(4.1)]{Braun:2009vc}, 
but in the both cases the eigenvalues are the same, $2S_1(N)$.

\begin{widetext}
\begin{center}
\begin{table*}[t]
	\begin{tabular}{| L | L | }
		\hline
		h(\tau) & E_N^{(h)} \\
		\hline 
		\widehat{\mathbb H}  & 2S_1 \\
		\hline
		1 		&  1/N/(N+1) \\
		\hline
		1-\tau & (-1)^N \left( 2S_{-2} + \zeta_2\right)   \\
		\hline
		\tfrac12 (1-\tau)\ln\left(\tfrac{1-\tau}{\tau}\right) &  (-1)^N \left( 2 S_{-2,1} - S_{-3} + \tfrac12 \zeta_3\right)\\
		\hline
		\tfrac12 \tfrac{1-\tau}{\tau}\ln(1-\tau) &  S_3 - \zeta_3 \\
		\hline
		\tfrac12 (1-\tau)\left( \mathrm{Li}_2(\tau) + \tfrac12 \ln^2(\tau)\right) &  (-1)^N \left(S_{-4} + \tfrac7{20}\zeta_2^2\right)   \\
		\hline
		\tfrac14 \tfrac{1-\tau}{\tau} \left(\mathrm{Li}_2(\tau) + \tfrac12\ln^2(\tau)\right) &  S_{1,3} - \tfrac12 S_4 - \zeta_3 S_1 + \tfrac3{10}\zeta_2^2    \\
		\hline 
		\tfrac12(1-\tau)\ln^2(1-\tau) &  (-1)^N\, 2\left(  4 S_{1,1,-2} -2 S_{2,-2} - 2 S_{1,-3} + S_{-4}  - \zeta_2 S_2 - \zeta_3 S_1 + 2 \zeta_2 S_{1,1} + \tfrac15 \zeta_2^2\right)    \\
		\hline 
	\end{tabular}
\caption{Invariant kernels and their eigenvalues. All harmonic sums have argument $N$.
\label{table:1}}
\end{table*}
\end{center}
\end{widetext}

\bibliographystyle{apsrev}
\bibliography{references}

\begin{thebibliography}{49}
\expandafter\ifx\csname natexlab\endcsname\relax\def\natexlab#1{#1}\fi
\expandafter\ifx\csname bibnamefont\endcsname\relax
  \def\bibnamefont#1{#1}\fi
\expandafter\ifx\csname bibfnamefont\endcsname\relax
  \def\bibfnamefont#1{#1}\fi
\expandafter\ifx\csname citenamefont\endcsname\relax
  \def\citenamefont#1{#1}\fi
\expandafter\ifx\csname url\endcsname\relax
  \def\url#1{\texttt{#1}}\fi
\expandafter\ifx\csname urlprefix\endcsname\relax\def\urlprefix{URL }\fi
\providecommand{\bibinfo}[2]{#2}
\providecommand{\eprint}[2][]{\url{#2}}

\bibitem[{\citenamefont{M{\"u}ller et~al.}(1994)\citenamefont{M{\"u}ller,
  Robaschik, Geyer, Dittes, and Ho{\v{r}}ej{\v{s}}i}}]{Muller:1994ses}
\bibinfo{author}{\bibfnamefont{D.}~\bibnamefont{M{\"u}ller}},
  \bibinfo{author}{\bibfnamefont{D.}~\bibnamefont{Robaschik}},
  \bibinfo{author}{\bibfnamefont{B.}~\bibnamefont{Geyer}},
  \bibinfo{author}{\bibfnamefont{F.~M.} \bibnamefont{Dittes}},
  \bibnamefont{and}
  \bibinfo{author}{\bibfnamefont{J.}~\bibnamefont{Ho{\v{r}}ej{\v{s}}i}},
  \bibinfo{journal}{Fortsch. Phys.} \textbf{\bibinfo{volume}{42}},
  \bibinfo{pages}{101} (\bibinfo{year}{1994}), \eprint{hep-ph/9812448}.

\bibitem[{\citenamefont{Accardi et~al.}(2024)}]{Accardi:2023chb}
\bibinfo{author}{\bibfnamefont{A.}~\bibnamefont{Accardi}} \bibnamefont{et~al.},
  \bibinfo{journal}{Eur. Phys. J. A} \textbf{\bibinfo{volume}{60}},
  \bibinfo{pages}{173} (\bibinfo{year}{2024}), \eprint{2306.09360}.

\bibitem[{\citenamefont{Accardi et~al.}(2016)}]{Accardi:2012qut}
\bibinfo{author}{\bibfnamefont{A.}~\bibnamefont{Accardi}} \bibnamefont{et~al.},
  \bibinfo{journal}{Eur. Phys. J. A} \textbf{\bibinfo{volume}{52}},
  \bibinfo{pages}{268} (\bibinfo{year}{2016}), \eprint{1212.1701}.

\bibitem[{\citenamefont{Anderle et~al.}(2021)}]{Anderle:2021wcy}
\bibinfo{author}{\bibfnamefont{D.~P.} \bibnamefont{Anderle}}
  \bibnamefont{et~al.}, \bibinfo{journal}{Front. Phys. (Beijing)}
  \textbf{\bibinfo{volume}{16}}, \bibinfo{pages}{64701} (\bibinfo{year}{2021}),
  \eprint{2102.09222}.

\bibitem[{\citenamefont{Ji}(1997)}]{Ji:1996ek}
\bibinfo{author}{\bibfnamefont{X.-D.} \bibnamefont{Ji}},
  \bibinfo{journal}{Phys. Rev. Lett.} \textbf{\bibinfo{volume}{78}},
  \bibinfo{pages}{610} (\bibinfo{year}{1997}), \eprint{hep-ph/9603249}.

\bibitem[{\citenamefont{Radyushkin}(1996)}]{Radyushkin:1996nd}
\bibinfo{author}{\bibfnamefont{A.~V.} \bibnamefont{Radyushkin}},
  \bibinfo{journal}{Phys. Lett. B} \textbf{\bibinfo{volume}{380}},
  \bibinfo{pages}{417} (\bibinfo{year}{1996}), \eprint{hep-ph/9604317}.

\bibitem[{\citenamefont{Radyushkin}(1997)}]{Radyushkin:1997ki}
\bibinfo{author}{\bibfnamefont{A.~V.} \bibnamefont{Radyushkin}},
  \bibinfo{journal}{Phys. Rev. D} \textbf{\bibinfo{volume}{56}},
  \bibinfo{pages}{5524} (\bibinfo{year}{1997}), \eprint{hep-ph/9704207}.

\bibitem[{\citenamefont{Ji and Osborne}(1998{\natexlab{a}})}]{Ji:1998xh}
\bibinfo{author}{\bibfnamefont{X.-D.} \bibnamefont{Ji}} \bibnamefont{and}
  \bibinfo{author}{\bibfnamefont{J.}~\bibnamefont{Osborne}},
  \bibinfo{journal}{Phys. Rev. D} \textbf{\bibinfo{volume}{58}},
  \bibinfo{pages}{094018} (\bibinfo{year}{1998}{\natexlab{a}}),
  \eprint{hep-ph/9801260}.

\bibitem[{\citenamefont{Collins and Freund}(1999)}]{Collins:1998be}
\bibinfo{author}{\bibfnamefont{J.~C.} \bibnamefont{Collins}} \bibnamefont{and}
  \bibinfo{author}{\bibfnamefont{A.}~\bibnamefont{Freund}},
  \bibinfo{journal}{Phys. Rev. D} \textbf{\bibinfo{volume}{59}},
  \bibinfo{pages}{074009} (\bibinfo{year}{1999}), \eprint{hep-ph/9801262}.

\bibitem[{\citenamefont{Braun et~al.}(2020)\citenamefont{Braun, Manashov, Moch,
  and Schoenleber}}]{Braun:2020yib}
\bibinfo{author}{\bibfnamefont{V.~M.} \bibnamefont{Braun}},
  \bibinfo{author}{\bibfnamefont{A.~N.} \bibnamefont{Manashov}},
  \bibinfo{author}{\bibfnamefont{S.}~\bibnamefont{Moch}}, \bibnamefont{and}
  \bibinfo{author}{\bibfnamefont{J.}~\bibnamefont{Schoenleber}},
  \bibinfo{journal}{JHEP} \textbf{\bibinfo{volume}{09}}, \bibinfo{pages}{117}
  (\bibinfo{year}{2020}), \bibinfo{note}{[Erratum: JHEP 02, 115 (2022)]},
  \eprint{2007.06348}.

\bibitem[{\citenamefont{Braun et~al.}(2021)\citenamefont{Braun, Manashov, Moch,
  and Schoenleber}}]{Braun:2021grd}
\bibinfo{author}{\bibfnamefont{V.~M.} \bibnamefont{Braun}},
  \bibinfo{author}{\bibfnamefont{A.~N.} \bibnamefont{Manashov}},
  \bibinfo{author}{\bibfnamefont{S.}~\bibnamefont{Moch}}, \bibnamefont{and}
  \bibinfo{author}{\bibfnamefont{J.}~\bibnamefont{Schoenleber}},
  \bibinfo{journal}{Phys. Rev. D} \textbf{\bibinfo{volume}{104}},
  \bibinfo{pages}{094007} (\bibinfo{year}{2021}), \eprint{2106.01437}.

\bibitem[{\citenamefont{Gao et~al.}(2022)\citenamefont{Gao, Huber, Ji, and
  Wang}}]{Gao:2021iqq}
\bibinfo{author}{\bibfnamefont{J.}~\bibnamefont{Gao}},
  \bibinfo{author}{\bibfnamefont{T.}~\bibnamefont{Huber}},
  \bibinfo{author}{\bibfnamefont{Y.}~\bibnamefont{Ji}}, \bibnamefont{and}
  \bibinfo{author}{\bibfnamefont{Y.-M.} \bibnamefont{Wang}},
  \bibinfo{journal}{Phys. Rev. Lett.} \textbf{\bibinfo{volume}{128}},
  \bibinfo{pages}{062003} (\bibinfo{year}{2022}), \eprint{2106.01390}.

\bibitem[{\citenamefont{Braun et~al.}(2022)\citenamefont{Braun, Ji, and
  Schoenleber}}]{Braun:2022bpn}
\bibinfo{author}{\bibfnamefont{V.~M.} \bibnamefont{Braun}},
  \bibinfo{author}{\bibfnamefont{Y.}~\bibnamefont{Ji}}, \bibnamefont{and}
  \bibinfo{author}{\bibfnamefont{J.}~\bibnamefont{Schoenleber}},
  \bibinfo{journal}{Phys. Rev. Lett.} \textbf{\bibinfo{volume}{129}},
  \bibinfo{pages}{172001} (\bibinfo{year}{2022}), \eprint{2207.06818}.

\bibitem[{\citenamefont{Ji and Schoenleber}(2024)}]{Ji:2023xzk}
\bibinfo{author}{\bibfnamefont{Y.}~\bibnamefont{Ji}} \bibnamefont{and}
  \bibinfo{author}{\bibfnamefont{J.}~\bibnamefont{Schoenleber}},
  \bibinfo{journal}{JHEP} \textbf{\bibinfo{volume}{01}}, \bibinfo{pages}{053}
  (\bibinfo{year}{2024}), \eprint{2310.05724}.

\bibitem[{\citenamefont{Braun et~al.}(2017)\citenamefont{Braun, Manashov, Moch,
  and Strohmaier}}]{Braun:2017cih}
\bibinfo{author}{\bibfnamefont{V.~M.} \bibnamefont{Braun}},
  \bibinfo{author}{\bibfnamefont{A.~N.} \bibnamefont{Manashov}},
  \bibinfo{author}{\bibfnamefont{S.}~\bibnamefont{Moch}}, \bibnamefont{and}
  \bibinfo{author}{\bibfnamefont{M.}~\bibnamefont{Strohmaier}},
  \bibinfo{journal}{JHEP} \textbf{\bibinfo{volume}{06}}, \bibinfo{pages}{037}
  (\bibinfo{year}{2017}), \eprint{1703.09532}.

\bibitem[{\citenamefont{Ji et~al.}(2023)\citenamefont{Ji, Manashov, and
  Moch}}]{Ji:2023eni}
\bibinfo{author}{\bibfnamefont{Y.}~\bibnamefont{Ji}},
  \bibinfo{author}{\bibfnamefont{A.}~\bibnamefont{Manashov}}, \bibnamefont{and}
  \bibinfo{author}{\bibfnamefont{S.-O.} \bibnamefont{Moch}},
  \bibinfo{journal}{Phys. Rev. D} \textbf{\bibinfo{volume}{108}},
  \bibinfo{pages}{054009} (\bibinfo{year}{2023}), \eprint{2307.01763}.

\bibitem[{\citenamefont{Manashov et~al.}(2005)\citenamefont{Manashov, Kirch,
  and Schafer}}]{Manashov:2005xp}
\bibinfo{author}{\bibfnamefont{A.}~\bibnamefont{Manashov}},
  \bibinfo{author}{\bibfnamefont{M.}~\bibnamefont{Kirch}}, \bibnamefont{and}
  \bibinfo{author}{\bibfnamefont{A.}~\bibnamefont{Schafer}},
  \bibinfo{journal}{Phys. Rev. Lett.} \textbf{\bibinfo{volume}{95}},
  \bibinfo{pages}{012002} (\bibinfo{year}{2005}), \eprint{hep-ph/0503109}.

\bibitem[{\citenamefont{Mueller and Schafer}(2006)}]{Mueller:2005ed}
\bibinfo{author}{\bibfnamefont{D.}~\bibnamefont{Mueller}} \bibnamefont{and}
  \bibinfo{author}{\bibfnamefont{A.}~\bibnamefont{Schafer}},
  \bibinfo{journal}{Nucl. Phys. B} \textbf{\bibinfo{volume}{739}},
  \bibinfo{pages}{1} (\bibinfo{year}{2006}), \eprint{hep-ph/0509204}.

\bibitem[{\citenamefont{Kirch et~al.}(2005)\citenamefont{Kirch, Manashov, and
  Schafer}}]{Kirch:2005tt}
\bibinfo{author}{\bibfnamefont{M.}~\bibnamefont{Kirch}},
  \bibinfo{author}{\bibfnamefont{A.}~\bibnamefont{Manashov}}, \bibnamefont{and}
  \bibinfo{author}{\bibfnamefont{A.}~\bibnamefont{Schafer}},
  \bibinfo{journal}{Phys. Rev. D} \textbf{\bibinfo{volume}{72}},
  \bibinfo{pages}{114006} (\bibinfo{year}{2005}), \eprint{hep-ph/0509330}.

\bibitem[{\citenamefont{Kumericki et~al.}(2008)\citenamefont{Kumericki,
  Mueller, and Passek-Kumericki}}]{Kumericki:2007sa}
\bibinfo{author}{\bibfnamefont{K.}~\bibnamefont{Kumericki}},
  \bibinfo{author}{\bibfnamefont{D.}~\bibnamefont{Mueller}}, \bibnamefont{and}
  \bibinfo{author}{\bibfnamefont{K.}~\bibnamefont{Passek-Kumericki}},
  \bibinfo{journal}{Nucl. Phys. B} \textbf{\bibinfo{volume}{794}},
  \bibinfo{pages}{244} (\bibinfo{year}{2008}), \eprint{hep-ph/0703179}.

\bibitem[{\citenamefont{Kumeri{\v{c}}ki and Mueller}(2010)}]{Kumericki:2009uq}
\bibinfo{author}{\bibfnamefont{K.}~\bibnamefont{Kumeri{\v{c}}ki}}
  \bibnamefont{and} \bibinfo{author}{\bibfnamefont{D.}~\bibnamefont{Mueller}},
  \bibinfo{journal}{Nucl. Phys. B} \textbf{\bibinfo{volume}{841}},
  \bibinfo{pages}{1} (\bibinfo{year}{2010}), \eprint{0904.0458}.

\bibitem[{\citenamefont{Zhang and Ji}(2025)}]{Zhang:2024djl}
\bibinfo{author}{\bibfnamefont{H.-C.} \bibnamefont{Zhang}} \bibnamefont{and}
  \bibinfo{author}{\bibfnamefont{X.}~\bibnamefont{Ji}}, \bibinfo{journal}{Nucl.
  Phys. B} \textbf{\bibinfo{volume}{1010}}, \bibinfo{pages}{116762}
  (\bibinfo{year}{2025}), \eprint{2408.04133}.

\bibitem[{\citenamefont{Kumeri\v{c}ki}()}]{Gepard}
\bibinfo{author}{\bibfnamefont{K.}~\bibnamefont{Kumeri\v{c}ki}},
  \emph{\bibinfo{title}{\textit{Gepard: Tool for studying the 3D quark and
  gluon distributions in the nucleon}}},
  \bibinfo{howpublished}{\url{https://gepard.phy.hr}}.

\bibitem[{\citenamefont{M{\"u}ller et~al.}(2014)\citenamefont{M{\"u}ller,
  Lautenschlager, Passek-Kumericki, and Schaefer}}]{Muller:2013jur}
\bibinfo{author}{\bibfnamefont{D.}~\bibnamefont{M{\"u}ller}},
  \bibinfo{author}{\bibfnamefont{T.}~\bibnamefont{Lautenschlager}},
  \bibinfo{author}{\bibfnamefont{K.}~\bibnamefont{Passek-Kumericki}},
  \bibnamefont{and} \bibinfo{author}{\bibfnamefont{A.}~\bibnamefont{Schaefer}},
  \bibinfo{journal}{Nucl. Phys. B} \textbf{\bibinfo{volume}{884}},
  \bibinfo{pages}{438} (\bibinfo{year}{2014}), \eprint{1310.5394}.

\bibitem[{\citenamefont{{\v{C}}ui{\'c}
  et~al.}(2023)\citenamefont{{\v{C}}ui{\'c}, Duplan{\v{c}}i{\'c},
  Kumeri{\v{c}}ki, and Passek-K.}}]{Cuic:2023mki}
\bibinfo{author}{\bibfnamefont{M.}~\bibnamefont{{\v{C}}ui{\'c}}},
  \bibinfo{author}{\bibfnamefont{G.}~\bibnamefont{Duplan{\v{c}}i{\'c}}},
  \bibinfo{author}{\bibfnamefont{K.}~\bibnamefont{Kumeri{\v{c}}ki}},
  \bibnamefont{and}
  \bibinfo{author}{\bibfnamefont{K.}~\bibnamefont{Passek-K.}},
  \bibinfo{journal}{JHEP} \textbf{\bibinfo{volume}{12}}, \bibinfo{pages}{192}
  (\bibinfo{year}{2023}), \bibinfo{note}{[Erratum: JHEP 02, 225 (2024)]},
  \eprint{2310.13837}.

\bibitem[{\citenamefont{Guo et~al.}(2022{\natexlab{a}})\citenamefont{Guo, Ji,
  and Shiells}}]{Guo:2022upw}
\bibinfo{author}{\bibfnamefont{Y.}~\bibnamefont{Guo}},
  \bibinfo{author}{\bibfnamefont{X.}~\bibnamefont{Ji}}, \bibnamefont{and}
  \bibinfo{author}{\bibfnamefont{K.}~\bibnamefont{Shiells}},
  \bibinfo{journal}{JHEP} \textbf{\bibinfo{volume}{09}}, \bibinfo{pages}{215}
  (\bibinfo{year}{2022}{\natexlab{a}}), \eprint{2207.05768}.

\bibitem[{\citenamefont{Guo et~al.}(2022{\natexlab{b}})}]{GUMP}
\bibinfo{author}{\bibfnamefont{Y.}~\bibnamefont{Guo}} \bibnamefont{et~al.},
  \emph{\bibinfo{title}{{GUMP GPD Global Analysis}}},
  \bibinfo{howpublished}{\url{https://github.com/yuxunguo/GUMP-Global-GPDs}}
  (\bibinfo{year}{2022}{\natexlab{b}}).

\bibitem[{\citenamefont{Guo et~al.}(2025{\natexlab{a}})\citenamefont{Guo, Ji,
  Santiago, Yang, and Zhang}}]{Guo:2024wxy}
\bibinfo{author}{\bibfnamefont{Y.}~\bibnamefont{Guo}},
  \bibinfo{author}{\bibfnamefont{X.}~\bibnamefont{Ji}},
  \bibinfo{author}{\bibfnamefont{M.~G.} \bibnamefont{Santiago}},
  \bibinfo{author}{\bibfnamefont{J.}~\bibnamefont{Yang}}, \bibnamefont{and}
  \bibinfo{author}{\bibfnamefont{H.-C.} \bibnamefont{Zhang}},
  \bibinfo{journal}{Phys. Rev. D} \textbf{\bibinfo{volume}{112}},
  \bibinfo{pages}{054036} (\bibinfo{year}{2025}{\natexlab{a}}),
  \eprint{2409.17231}.

\bibitem[{\citenamefont{Guo et~al.}(2025{\natexlab{b}})\citenamefont{Guo,
  Aslan, Ji, and Santiago}}]{Guo:2025muf}
\bibinfo{author}{\bibfnamefont{Y.}~\bibnamefont{Guo}},
  \bibinfo{author}{\bibfnamefont{F.~P.} \bibnamefont{Aslan}},
  \bibinfo{author}{\bibfnamefont{X.}~\bibnamefont{Ji}}, \bibnamefont{and}
  \bibinfo{author}{\bibfnamefont{M.~G.} \bibnamefont{Santiago}}
  (\bibinfo{year}{2025}{\natexlab{b}}), \eprint{2509.08037}.

\bibitem[{\citenamefont{Mamo and Zahed}(2024{\natexlab{a}})}]{Mamo:2024jwp}
\bibinfo{author}{\bibfnamefont{K.~A.} \bibnamefont{Mamo}} \bibnamefont{and}
  \bibinfo{author}{\bibfnamefont{I.}~\bibnamefont{Zahed}},
  \bibinfo{journal}{Phys. Rev. Lett.} \textbf{\bibinfo{volume}{133}},
  \bibinfo{pages}{241901} (\bibinfo{year}{2024}{\natexlab{a}}),
  \eprint{2411.04162}.

\bibitem[{\citenamefont{Mamo and Zahed}(2024{\natexlab{b}})}]{Mamo:2024vjh}
\bibinfo{author}{\bibfnamefont{K.~A.} \bibnamefont{Mamo}} \bibnamefont{and}
  \bibinfo{author}{\bibfnamefont{I.}~\bibnamefont{Zahed}},
  \bibinfo{journal}{Phys. Rev. D} \textbf{\bibinfo{volume}{110}},
  \bibinfo{pages}{114016} (\bibinfo{year}{2024}{\natexlab{b}}),
  \eprint{2404.13245}.

\bibitem[{\citenamefont{Hechenberger et~al.}(2025)\citenamefont{Hechenberger,
  Mamo, and Zahed}}]{Hechenberger:2025wnz}
\bibinfo{author}{\bibfnamefont{F.}~\bibnamefont{Hechenberger}},
  \bibinfo{author}{\bibfnamefont{K.~A.} \bibnamefont{Mamo}}, \bibnamefont{and}
  \bibinfo{author}{\bibfnamefont{I.}~\bibnamefont{Zahed}},
  \bibinfo{journal}{Phys. Rev. D} \textbf{\bibinfo{volume}{112}},
  \bibinfo{pages}{074018} (\bibinfo{year}{2025}), \eprint{2508.00817}.

\bibitem[{\citenamefont{Panzer}(2015)}]{Panzer:2014caa}
\bibinfo{author}{\bibfnamefont{E.}~\bibnamefont{Panzer}},
  \bibinfo{journal}{Comput. Phys. Commun.} \textbf{\bibinfo{volume}{188}},
  \bibinfo{pages}{148} (\bibinfo{year}{2015}), \eprint{1403.3385}.

\bibitem[{\citenamefont{Duhr and Dulat}(2019)}]{Duhr:2019tlz}
\bibinfo{author}{\bibfnamefont{C.}~\bibnamefont{Duhr}} \bibnamefont{and}
  \bibinfo{author}{\bibfnamefont{F.}~\bibnamefont{Dulat}},
  \bibinfo{journal}{JHEP} \textbf{\bibinfo{volume}{08}}, \bibinfo{pages}{135}
  (\bibinfo{year}{2019}), \eprint{1904.07279}.

\bibitem[{\citenamefont{Belitsky and Radyushkin}(2005)}]{Belitsky:2005qn}
\bibinfo{author}{\bibfnamefont{A.~V.} \bibnamefont{Belitsky}} \bibnamefont{and}
  \bibinfo{author}{\bibfnamefont{A.~V.} \bibnamefont{Radyushkin}},
  \bibinfo{journal}{Phys. Rept.} \textbf{\bibinfo{volume}{418}},
  \bibinfo{pages}{1} (\bibinfo{year}{2005}), \eprint{hep-ph/0504030}.

\bibitem[{\citenamefont{Mueller}(2006)}]{Mueller:2005nz}
\bibinfo{author}{\bibfnamefont{D.}~\bibnamefont{Mueller}},
  \bibinfo{journal}{Phys. Lett. B} \textbf{\bibinfo{volume}{634}},
  \bibinfo{pages}{227} (\bibinfo{year}{2006}), \eprint{hep-ph/0510109}.

\bibitem[{\citenamefont{Kumericki et~al.}(2007)\citenamefont{Kumericki,
  Mueller, Passek-Kumericki, and Schafer}}]{Kumericki:2006xx}
\bibinfo{author}{\bibfnamefont{K.}~\bibnamefont{Kumericki}},
  \bibinfo{author}{\bibfnamefont{D.}~\bibnamefont{Mueller}},
  \bibinfo{author}{\bibfnamefont{K.}~\bibnamefont{Passek-Kumericki}},
  \bibnamefont{and} \bibinfo{author}{\bibfnamefont{A.}~\bibnamefont{Schafer}},
  \bibinfo{journal}{Phys. Lett. B} \textbf{\bibinfo{volume}{648}},
  \bibinfo{pages}{186} (\bibinfo{year}{2007}), \eprint{hep-ph/0605237}.

\bibitem[{\citenamefont{Ji and Osborne}(1998{\natexlab{b}})}]{Ji:1997nk}
\bibinfo{author}{\bibfnamefont{X.-D.} \bibnamefont{Ji}} \bibnamefont{and}
  \bibinfo{author}{\bibfnamefont{J.}~\bibnamefont{Osborne}},
  \bibinfo{journal}{Phys. Rev. D} \textbf{\bibinfo{volume}{57}},
  \bibinfo{pages}{1337} (\bibinfo{year}{1998}{\natexlab{b}}),
  \eprint{hep-ph/9707254}.

\bibitem[{\citenamefont{Mankiewicz et~al.}(1998)\citenamefont{Mankiewicz,
  Piller, Stein, Vanttinen, and Weigl}}]{Mankiewicz:1997bk}
\bibinfo{author}{\bibfnamefont{L.}~\bibnamefont{Mankiewicz}},
  \bibinfo{author}{\bibfnamefont{G.}~\bibnamefont{Piller}},
  \bibinfo{author}{\bibfnamefont{E.}~\bibnamefont{Stein}},
  \bibinfo{author}{\bibfnamefont{M.}~\bibnamefont{Vanttinen}},
  \bibnamefont{and} \bibinfo{author}{\bibfnamefont{T.}~\bibnamefont{Weigl}},
  \bibinfo{journal}{Phys. Lett. B} \textbf{\bibinfo{volume}{425}},
  \bibinfo{pages}{186} (\bibinfo{year}{1998}), \bibinfo{note}{[Erratum:
  Phys.Lett.B 461, 423--423 (1999)]}, \eprint{hep-ph/9712251}.

\bibitem[{\citenamefont{Belitsky and Mueller}(1998)}]{Belitsky:1997rh}
\bibinfo{author}{\bibfnamefont{A.~V.} \bibnamefont{Belitsky}} \bibnamefont{and}
  \bibinfo{author}{\bibfnamefont{D.}~\bibnamefont{Mueller}},
  \bibinfo{journal}{Phys. Lett. B} \textbf{\bibinfo{volume}{417}},
  \bibinfo{pages}{129} (\bibinfo{year}{1998}), \eprint{hep-ph/9709379}.

\bibitem[{\citenamefont{Hoodbhoy and Ji}(1998)}]{Hoodbhoy:1998vm}
\bibinfo{author}{\bibfnamefont{P.}~\bibnamefont{Hoodbhoy}} \bibnamefont{and}
  \bibinfo{author}{\bibfnamefont{X.-D.} \bibnamefont{Ji}},
  \bibinfo{journal}{Phys. Rev. D} \textbf{\bibinfo{volume}{58}},
  \bibinfo{pages}{054006} (\bibinfo{year}{1998}), \eprint{hep-ph/9801369}.

\bibitem[{\citenamefont{Belitsky and Mueller}(2000)}]{Belitsky:2000jk}
\bibinfo{author}{\bibfnamefont{A.~V.} \bibnamefont{Belitsky}} \bibnamefont{and}
  \bibinfo{author}{\bibfnamefont{D.}~\bibnamefont{Mueller}},
  \bibinfo{journal}{Phys. Lett. B} \textbf{\bibinfo{volume}{486}},
  \bibinfo{pages}{369} (\bibinfo{year}{2000}), \eprint{hep-ph/0005028}.

\bibitem[{\citenamefont{Remiddi and Vermaseren}(2000)}]{Remiddi:1999ew}
\bibinfo{author}{\bibfnamefont{E.}~\bibnamefont{Remiddi}} \bibnamefont{and}
  \bibinfo{author}{\bibfnamefont{J.~A.~M.} \bibnamefont{Vermaseren}},
  \bibinfo{journal}{Int. J. Mod. Phys. A} \textbf{\bibinfo{volume}{15}},
  \bibinfo{pages}{725} (\bibinfo{year}{2000}), \eprint{hep-ph/9905237}.

\bibitem[{\citenamefont{Melic et~al.}(2003)\citenamefont{Melic, Mueller, and
  Passek-Kumericki}}]{Melic:2002ij}
\bibinfo{author}{\bibfnamefont{B.}~\bibnamefont{Melic}},
  \bibinfo{author}{\bibfnamefont{D.}~\bibnamefont{Mueller}}, \bibnamefont{and}
  \bibinfo{author}{\bibfnamefont{K.}~\bibnamefont{Passek-Kumericki}},
  \bibinfo{journal}{Phys. Rev. D} \textbf{\bibinfo{volume}{68}},
  \bibinfo{pages}{014013} (\bibinfo{year}{2003}), \eprint{hep-ph/0212346}.

\bibitem[{\citenamefont{Braun et~al.}(2010)\citenamefont{Braun, Manashov, and
  Rohrwild}}]{Braun:2009vc}
\bibinfo{author}{\bibfnamefont{V.~M.} \bibnamefont{Braun}},
  \bibinfo{author}{\bibfnamefont{A.~N.} \bibnamefont{Manashov}},
  \bibnamefont{and} \bibinfo{author}{\bibfnamefont{J.}~\bibnamefont{Rohrwild}},
  \bibinfo{journal}{Nucl. Phys. B} \textbf{\bibinfo{volume}{826}},
  \bibinfo{pages}{235} (\bibinfo{year}{2010}), \eprint{0908.1684}.

\bibitem[{Sup()}]{Supplementary}
\emph{\bibinfo{title}{{See Supplementary Material
  http://link.aps.org/supplemental/10.1103/p5rk-497z for the Mathematica
  expressions.}}}

\bibitem[{\citenamefont{Dokshitzer et~al.}(2006)\citenamefont{Dokshitzer,
  Marchesini, and Salam}}]{Dokshitzer:2005bf}
\bibinfo{author}{\bibfnamefont{Y.~L.} \bibnamefont{Dokshitzer}},
  \bibinfo{author}{\bibfnamefont{G.}~\bibnamefont{Marchesini}},
  \bibnamefont{and} \bibinfo{author}{\bibfnamefont{G.~P.} \bibnamefont{Salam}},
  \bibinfo{journal}{Phys. Lett. B} \textbf{\bibinfo{volume}{634}},
  \bibinfo{pages}{504} (\bibinfo{year}{2006}), \eprint{hep-ph/0511302}.

\bibitem[{\citenamefont{Dokshitzer and Marchesini}(2007)}]{Dokshitzer:2006nm}
\bibinfo{author}{\bibfnamefont{Y.~L.} \bibnamefont{Dokshitzer}}
  \bibnamefont{and}
  \bibinfo{author}{\bibfnamefont{G.}~\bibnamefont{Marchesini}},
  \bibinfo{journal}{Phys. Lett. B} \textbf{\bibinfo{volume}{646}},
  \bibinfo{pages}{189} (\bibinfo{year}{2007}), \eprint{hep-th/0612248}.

\bibitem[{\citenamefont{Beccaria et~al.}(2009)\citenamefont{Beccaria, Forini,
  Lukowski, and Zieme}}]{Beccaria:2009eq}
\bibinfo{author}{\bibfnamefont{M.}~\bibnamefont{Beccaria}},
  \bibinfo{author}{\bibfnamefont{V.}~\bibnamefont{Forini}},
  \bibinfo{author}{\bibfnamefont{T.}~\bibnamefont{Lukowski}}, \bibnamefont{and}
  \bibinfo{author}{\bibfnamefont{S.}~\bibnamefont{Zieme}},
  \bibinfo{journal}{JHEP} \textbf{\bibinfo{volume}{03}}, \bibinfo{pages}{129}
  (\bibinfo{year}{2009}), \eprint{0901.4864}.

\end{thebibliography}

\end{document}